\documentstyle[pre,aps,multicol,epsf]{revtex}  
\def\B.#1{{\bbox{#1}}}  
\def\C.#1{{\cal{#1}}}

\def\i{{\rm i}}
\def\d{{\rm d}}

\def\be{\begin{equation}} 
\def\ee{\end{equation}} 
 
\def\alpha{k} 
\def\Re{{\rm Re}} 

\begin{document} 
\title{{\rm PRL} \hfill {\sl Submitted}\\ Retardation of the Onset of
Turbulence by Minor Viscosity Contrasts} \author{Rama
Govindarajan$^{\dag,*}$, Victor S. L'vov$^*$ and Itamar Procaccia$^*$}
\address{$^*$ Dept. of Chemical Physics, The Weizmann Institute of
Science, Rehovot 76100, Israel \\ $^\dag $ Fluid Dynamics Unit,
Jawaharlal Nehru Centre for Advanced Scientific Research, Jakkur,
Bangalore 560064, India.} \maketitle
 \begin{abstract}  
Motivated by the large effect of turbulent drag reduction by minute
concentrations of polymers we study the effects of minor viscosity
contrasts on the stability of hydrodynamic flows. The key player is a
localized region where the energy of fluctuations is produced by
interactions with the mean flow (the ``critical layer"). We show that
a layer of weakly space-dependent viscosity placed near the critical
layer can have very large stabilizing effect on hydrodynamic
fluctuations, retarding significantly the onset of turbulence. The
effect is {\em not} due to a modified dissipation (as is assumed in
theories of drag reduction), but due to reduced energy intake from the
mean flow to the fluctuations.  We propose that similar physics act in
turbulent drag reduction.
\end{abstract}
 
\begin{multicols}{2} 
 
The addition of small amounts of polymers to hydrodynamic systems
produces dramatic effects on phenomena such as the transition to
turbulence, vortex formation and turbulent transport \cite{95NH}. The
phenomenon that attracted most attention was, for obvious reasons, the
reduction of friction drag by up to 80\% when a very small
concentration of long-chain polymers were added to turbulent flows
\cite{69Lum,00SW}. In spite of the fact that the phenomenon is robust
and the effect huge, there exists no accepted theory that can claim
quantitative agreement with the experimental facts.  Moreover, it
appears that there is no mechanistic explanation.  In the current
theory that is due to de Gennes \cite{86TG,90Gennes} one expects the
Kolmogorov cascade to be terminated at scales larger than Kolmogorov
scale, leading somehow to an increased buffer layer thickness and
reduced drag, but how this happens and what is the fate of the
turbulent energy is not being made clear.

In this Letter we propose that the crucial issue is in the {\em
production} of energy of hydrodynamic fluctuations by their
interaction with the mean flow.  For the sake of concreteness we
examine a simple laminar flow and its loss of stability, and show how
small viscosity contrasts lead to an order of magnitude retardation in
the onset of instability of ``dangerous" disturbances. In this model
everything is explicitly calculable, and we demonstrate that nothing
special happens to the dissipation. Rather, it is the energy
production that is dramatically reduced, giving rise to a large effect
for a small cause. At the end of this Letter we argue that similar
physics may very well be at the heart of turbulent drag reduction, but
we stress that the phenomenon discussed below is interesting by itself
and well warrants an experimental confirmation.

It is well known that parallel Poiseuille flow loses its stability at
some threshold Reynolds number Re=Re$_{\rm th}$ (close to 5772).  It
is also well known that the instability is ``convective", with the
most unstable mode having a phase velocity $c$. Analytically it has
the form
\begin{equation} 
  \label{eq:inst}  
   \hat \phi(x,y,t) = \case{1}{2}\{\phi(y) 
\exp [\i \alpha(x-c \, t)] + 
   \mbox{c.c.} \} \exp(\gamma t) \ , 
\end{equation}
where $\hat \phi(x,y,t)$ is the disturbance streamfunction and
$\phi(y)$ is the complex envelope of $\hat \phi(x,y,t)$.  We have
chosen
\begin{figure} 
\epsfxsize=7.5cm \epsfbox{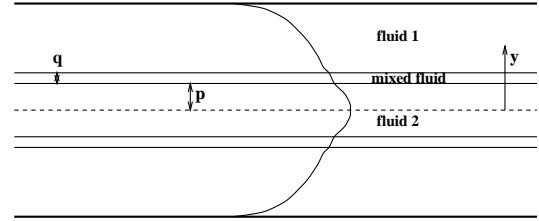}
\vskip 0.3cm  
\caption{Schematic of the flow: the fluid near the walls has a 
viscosity $\mu_1$, and that flowing at the center is of
viscosity $\mu_2$. In the mixed layer
(of width $q$) the viscosity varies gradually between $\mu_1$ and $\mu_2$. The
parameter $p$ controls the position  of the mixed layer.}
 \label{f:scheme} 
\end{figure} 
\noindent
 $x$ and $\alpha$ as the streamwise coordinate and wavenumber of
the disturbance, $c$ as the phase speed and $t$ as time. $\gamma$ is
the increment of instability.  What is not usually emphasized is that
the main interactions leading to the loss of stability occur in a
sharply defined region in the channel, i.e. at a layer whose distance
from the wall is such that the phase velocity $c$ is comparable to the
velocity of the mean flow. We refer to this layer as the ``critical"
layer.  It is thus worthwhile to examine the effect on the stability
of Poiseuille flow of a viscosity gradient placed in the vicinity of
the critical layer. Following \cite{rama} we examine a channel flow of
two fluids with different viscosities $\mu_1$ and $\mu_2$, see
Fig. \ref{f:scheme}.

Observing that the inferred effective viscosity in polymer drag
reduction increases towards the center by about 30\% over about a 1/3
of the half-channel \cite{97SBH}, we choose $\mu_2=1$ and $m\equiv
\mu_1/ \mu_2=0.9$, with all the viscosity difference of 0.1
concentrated in a ``mixing" layer of width 0.1, leading to comparable
viscosity gradients. The observation that we want to focus on is shown
in Fig. 2: the threshold Reynolds number for the loss of stability of
the mode as in Eq. (\ref{eq:inst}) depends crucially on the position
of the mixing layer. When the latter hits the critical layer the
threshold Reynolds number for the loss of stability reaches as much as
88000. In other words, one can increase the threshold of instability
{\em for a given mode} 15 times, and by making the mixing layer
thinner one can reach even higher threshold Reynolds values.
\begin{figure}
\epsfxsize=8.3 cm
\epsfbox{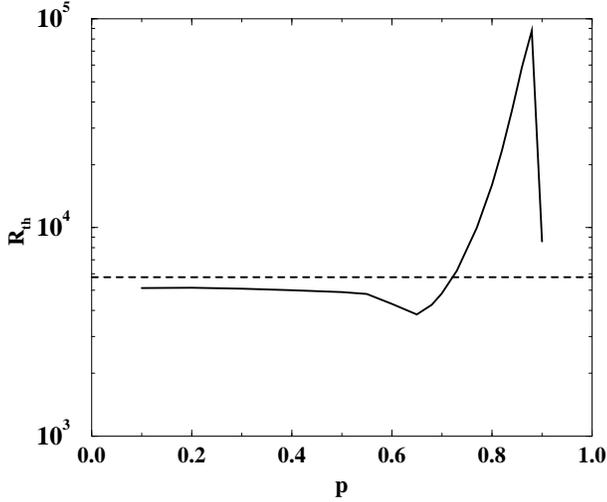}
\caption{The dependence of the threshold Reynolds number on the
position of the viscosity stratified layer for $m=0.9$. The dashed
line pertains to the neat fluid. Note the huge increase in $R_{\rm
th}$ within a small range. This occurs when the stratified layer
overlaps the critical layer.}
\end{figure}
In this Letter we analyze the physical origin of this huge sensitivity
of the flow stability to the profile of the viscosity.  The stability
of this flow is governed by the modified Orr-Sommerfeld equation
\cite{white}
\begin{eqnarray}\nonumber  
&&  \i\alpha \left[\left(\phi''-\alpha^2\phi\right)(U-c-\i \gamma)  
- U''\phi \right] = 
{1\over \Re}\bigg[\mu \phi^{\rm (4)}+ 2 \mu' \phi''' \\ \label{modOS} 
&& +\left(\mu'' - 2 \alpha^2 \mu \right)\phi'' - 2 \alpha^2 \mu' 
\phi' + \left(\alpha^2 \mu'' + \alpha^4 \mu\right) \phi\bigg]\,,  
\end{eqnarray} 
in which $\phi$, $U$ and $\mu$ are functions of $y$. The boundary
conditions are $\phi(\pm 1) = \phi' (\pm 1) = 0$.  All quantities have
been non-dimensionalised using the half-width $H$ of the channel and
the centerline velocity $U_0$ as the length and velocity scales
respectively.  The Reynolds number is defined as $\Re \equiv \rho U_0
H /\mu_2$, where $\rho$ is the density (equal for the two fluids).
The primes stand for derivative with respect to $y$.  At $y=0$, we use
the even symmetry conditions $\phi(0)=1$ and $\phi'(0)=0$, as the even
mode is always more unstable than the odd.

Since the flow is symmetric with respect to the channel centerline, we
restrict our attention to the upper half-channel. Fluid 2 occupies the
region $0 \le y \le p$. Fluid 1 lies between $p+q \le y \le 1$. The
region $p \le y \le p+q$ contains mixed fluid. The viscosity is
described by a steady function of $y$, scaled by the inner fluid
viscosity $\mu_2$:
\begin{eqnarray}\label{muin}
   \mu(y) &=& 1 \,,\quad  \mbox{for} \quad 0 \le y \le p\,,\\
\label{vis5}
\mu(y) &=& 1 + (m-1)\, \xi^3\left[10 - 15\, \xi + 6 \xi^2
\right], \ 0 \le \xi \le 1\,, \\
\label{muout}
   \mu(y) &=& m \,,\quad  \mbox{for} \quad p+q \le y \le 1\,,\\
\end{eqnarray}
Here $\xi\equiv (y-p)/q$ is the mixed layer coordinate.  We have
assumed a 5th-order polynomial profile for the viscosity in the mixed
layer, whose coefficients maintain the 
\begin{figure} 
\epsfxsize=7.5cm  \epsfbox{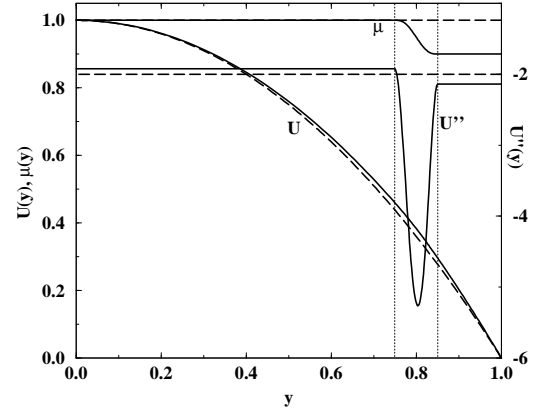} 
\caption{Profiles of the normalized viscosity $\mu(y)$ 
and normalized velocity $U(y)$ and the second derivative $U^{''}(y)$
for $m=0.9$ (solid lines) and $m=1.0$ (dashed lines).  The mixed layer
is between the vertical dashed lines. }
\label{f:prof} 
\end{figure} 
\noindent
viscosity and its first two derivatives continuous across the mixed
layer.  The exact form of the profile is unimportant. For a plot of
the profile $m=0.9$, see Fig.~\ref{f:prof}.

The basic flow $U(y)$ is obtained by requiring the velocity and all
relevant derivatives to be continuous at the edges of the mixed layer:
\begin{eqnarray}\label{ui}
&   U(y) = 1-Gy^2/2\,, \quad &\mbox{for} \quad y \le p\,,\\
&U(y) = U(p) - G\int_p^y{y\over \mu} \d y\,,\quad   &\mbox{for}
\quad p \le y \le p+ q,
\label{um}\\
& U(y) = G\left(1-y^2\right)/2m, \quad  &\mbox{for}
\quad y \ge p+q\ .\label{uo}
\end{eqnarray}
Here $G$ is the streamwise pressure gradient.  

It can be seen, comparing the mean profile $U(y)$ to that of the neat
fluid (cf. Fig. \ref{f:prof}), that nothing dramatic happens to this
profile even when the mixing layer is chosen to overlap a typical
critical layer. Accordingly we need to look for the origin of the
large effect of Fig. 2 in the energetics of the disturbances. To do
so, recall that the streamwise and normal components of the
disturbance velocity $\hat u(x,y,t)$ and $\hat v(x,y,t)$ may be
expressed via streamfunction as usual:
\begin{equation} 
  \label{velcomp} 
\hat u(x,y,t) = \partial\hat \phi/\partial y \quad {\rm and} \quad \hat 
v(x,y,t) = -\partial \hat \phi/\partial x \ . 
\end{equation} 
These functions may be written in terms of complex envelopes similar 
to Eq.~(\ref{eq:inst}): 
\begin{eqnarray} 
  \label{eq:vel-env} 
  \hat u(x,y,t) &=& \case{1}{2}\big\{u(y) \exp\left[\i\alpha(x-c\,
  t)\right] + \mbox{c.c.}\big\} \exp(\gamma t) \,,\\ \nonumber \hat
  v(x,y,t) &=& \case{1}{2}\big\{v(y) \exp\left[\i\alpha(x-c\,
  t)\right] + \mbox{c.c.}\big\} \exp(\gamma t) \ .
\end{eqnarray} 
The pressure disturbance $\hat p$ is defined similarly. 

Define now a disturbance of the density of the kinetic energy 
\begin{equation}\label{kinen} 
  \hat E(x,y,t) = \case{1}{2}\left[ \hat u(x,y,t)^2 + \hat 
  v(x,y,t)^2\right] 
\end{equation} 
we can express the mean (over $x$) density of the kinetic energy as
follows:
\begin{eqnarray}\label{averen} 
  E(y,t)&\equiv& \left< \hat E(x,y,t)\right>_x =\C.E (y)\exp\,(2 
\gamma t)\,,
\\ \nonumber
\C.E (y)&=&\case{1}{4}\left( |u(y)|^2 + |v(y)|^2 \right)\ . 
  \end{eqnarray}

The physics of our phenomenon will be discussed in terms of the
balance equation for the averaged disturbance kinetic energy. Starting
from the linearized Navier-Stokes equations for $\hat u$ and $\hat v$,
dotting it with the disturbance velocity vector, averaging over one
cycle in $x$ and using Eqs. (\ref{eq:vel-env})-(\ref{averen}) leads to
\be 2 \gamma \,\C.E (y) = \nabla \cdot J(y) + W_+(y) - W_-(y) \,,
\label{enbal} 
\ee where the energy flux $J(y)$ in the $y$ direction, rates of energy
production (by the mean flow) $W_+(y)$ and energy dissipation (by the
viscosity) $W_-(y)$ are given by
\begin{eqnarray}\label{current} 
  J(y) &\equiv&  {\left[u(y) p^*(y)+\mbox{c.c.}\right] 
\over 4 \rho} + {1 \over \Re} \mu (y) \nabla \C.E(y)\,,\\
W_+(y) &\equiv&  - {1 \over 4} U'(y)\left[u(y)
 v^*(y) +\mbox{c.c.}  \right]\,,
\label{prod} \\
W_-(y) &\equiv& {\mu(y) \over \Re} \left\{2 \alpha^2 \C.E(y)
+ {1 \over 2}\left[|u'(y)|^2+|v'(y)|^2\right]\right\}\ .
\label{diss}
\end{eqnarray}
The superscript $*$ denotes the complex conjugate. To plot these
functions we need to solve Eq. (\ref{modOS}) as an eigenvalue problem,
to obtain $c$, $\gamma$, and $\phi(y)$ at given Re and $k$. The value
of $c$ determines the position of the critical layer.

It is convenient to compute and compare the space averaged production
and dissipation terms $\Gamma_+$ and $\Gamma_-$ defined by:
\begin{equation}\label{totpd}
  \Gamma_\pm
 \equiv\int_0^1 W_\pm(y) \d y \ \Big/ \int_0^1 \C.E (y) \d y
 \ .
\end{equation}
The local production of energy can be positive or negative, indicative
of energy transfer from the mean flow to the disturbance and vice-versa
respectively. The production in one region (where $ W_+(y)>0$) can be
partly canceled out by a ``counter-production'' in other region (where
$ W_+(y)<0$). 

\begin{figure} 
\epsfxsize=7.5cm \epsfbox{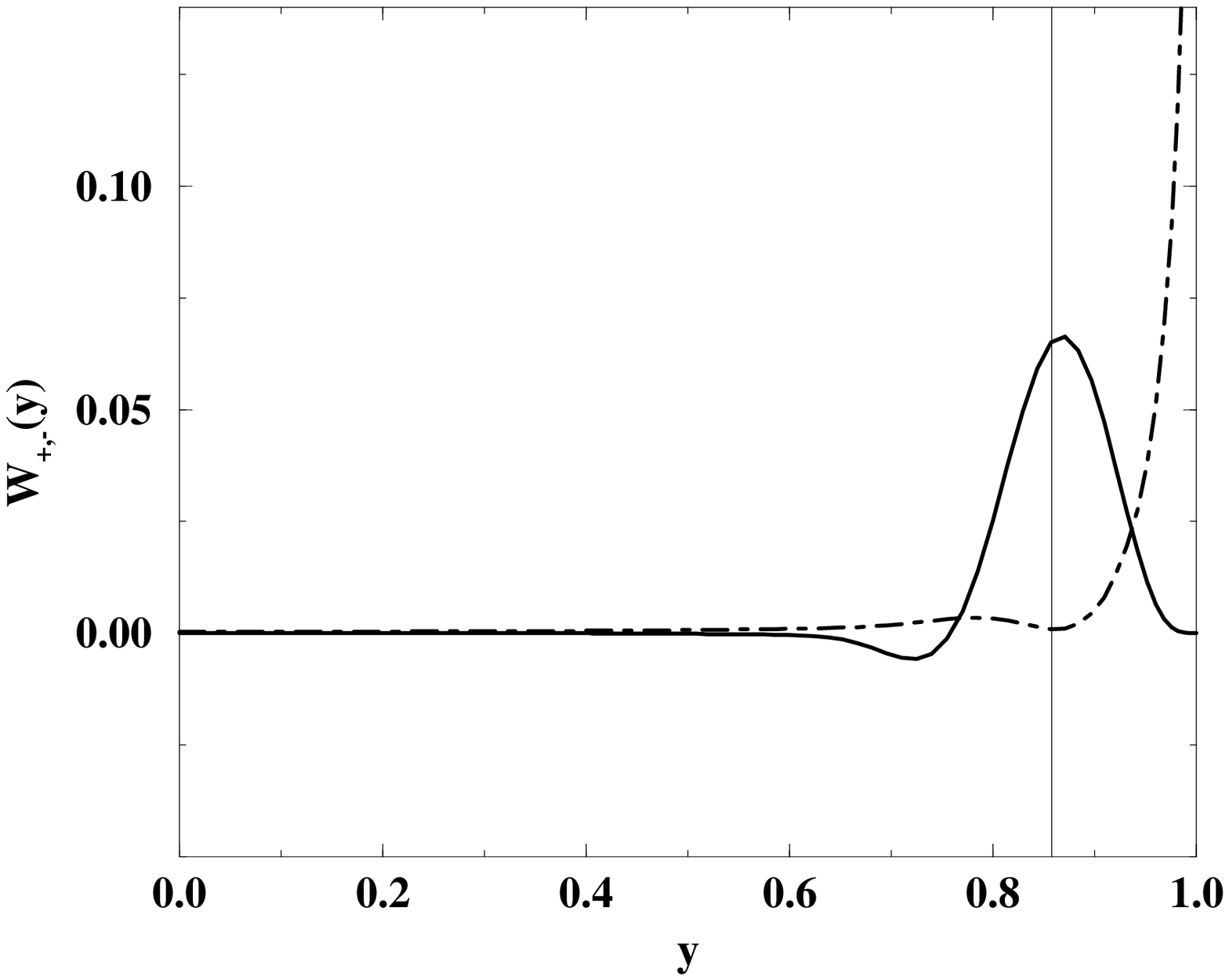}  
\vskip -1cm
\epsfxsize=7.5cm \epsfbox{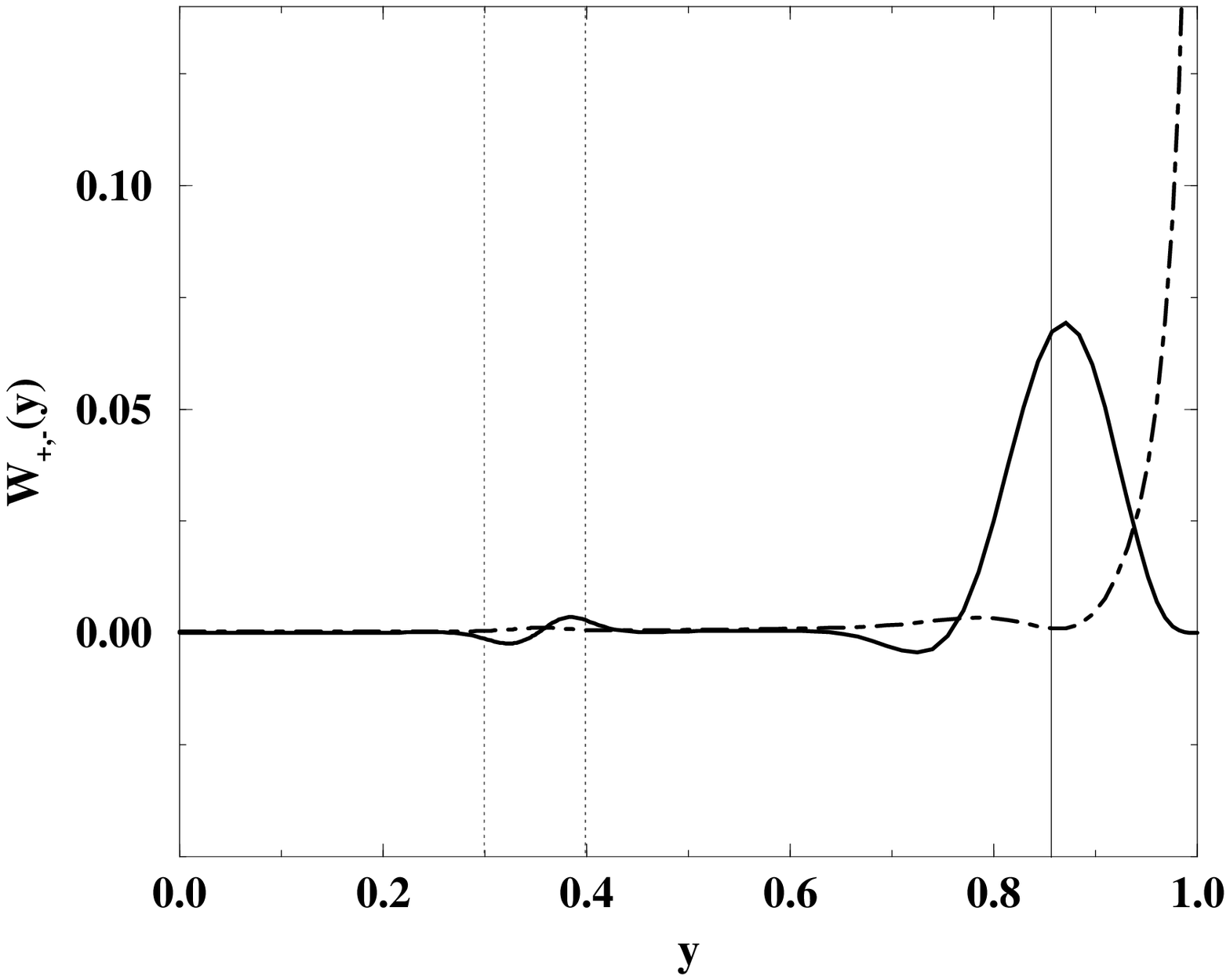}  
\caption{Energy balance: production $W_+(y)$, solid line; 
dissipation $W_-(y)$, dot-dashed line, $\Re=5772$. Top: $m=1$,
$\Gamma_+=\Gamma_-=0.0148$. Bottom: $m=0.9$, $p=0.3$,
$\Gamma_+=0.0158$, $\Gamma_-=0.0148$. In this and all the subsequent
figures the solid vertical lines show the location of the critical
lines, whereas the region between the dotted lines is the mixed
layer.}
 \label{f:far} 
\end{figure}

The use of these measures can be exemplified with the neat fluid
($m=1.0$ here). The laminar flow displays its first linear instability at
a threshold Reynolds number of $Re_{\rm th}=5772$, which means that
the total production $\Gamma_+$ across the layer becomes equal to the 
total dissipation $\Gamma_-$ at this value of $\Re$. Examining Fig. 
\ref{f:far} we can see that the disturbance kinetic energy is produced 
predominantly within the 
critical layer, where the basic flow velocity is close to the phase 
speed of the disturbance, while most of the dissipation is in the wall 
layer. 
The balance is not changed significantly when the viscosity ratio is
changed to $0.9$ so long as the mixed layer is not close to the
critical layer. There is a small region of production and one of
counter-production within the mixed layer, whose effects cancel out,
leaving the system close to marginal stability.

\begin{figure} 
\epsfxsize=7.5cm \epsfbox{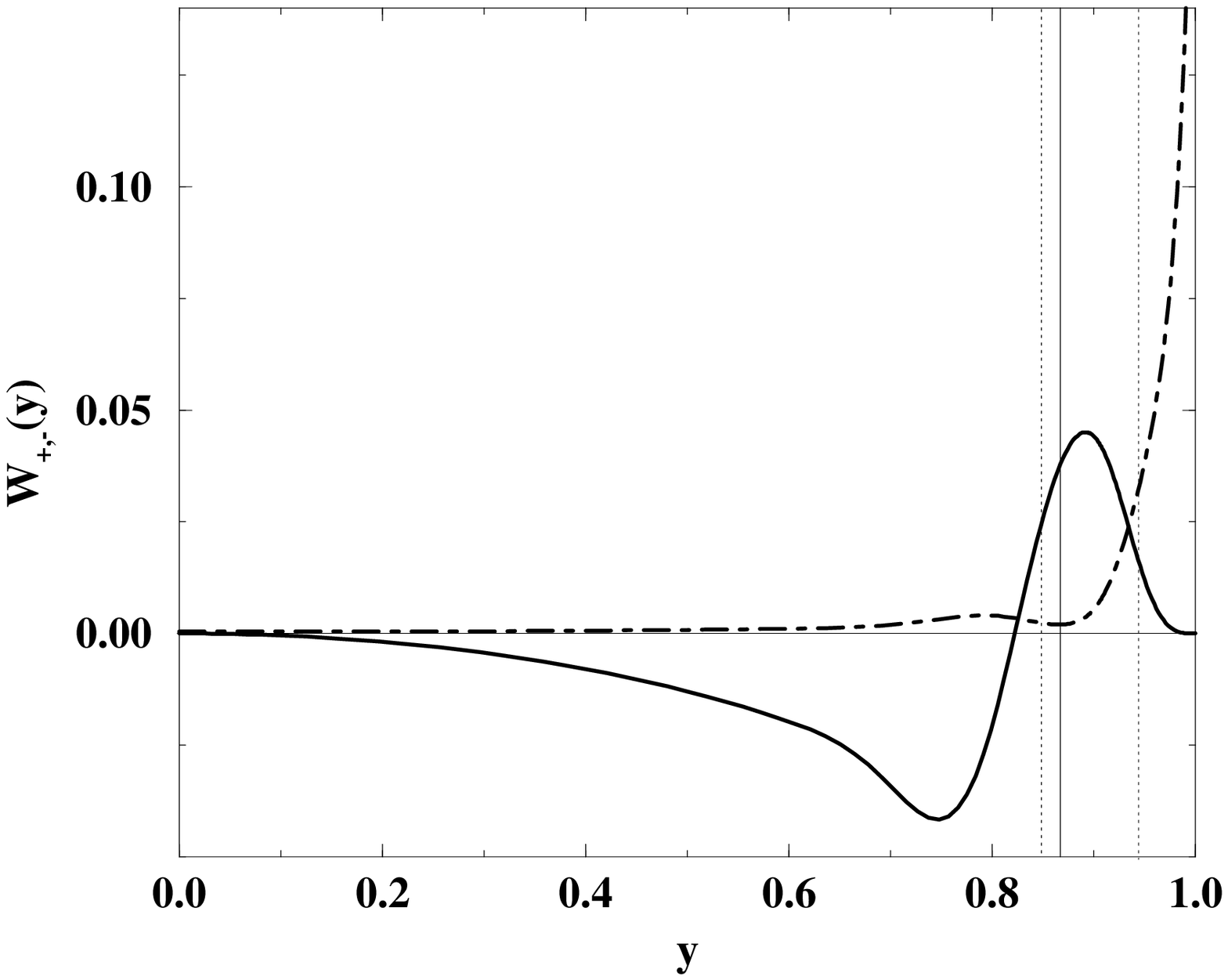}

\epsfxsize=7.5cm \epsfbox{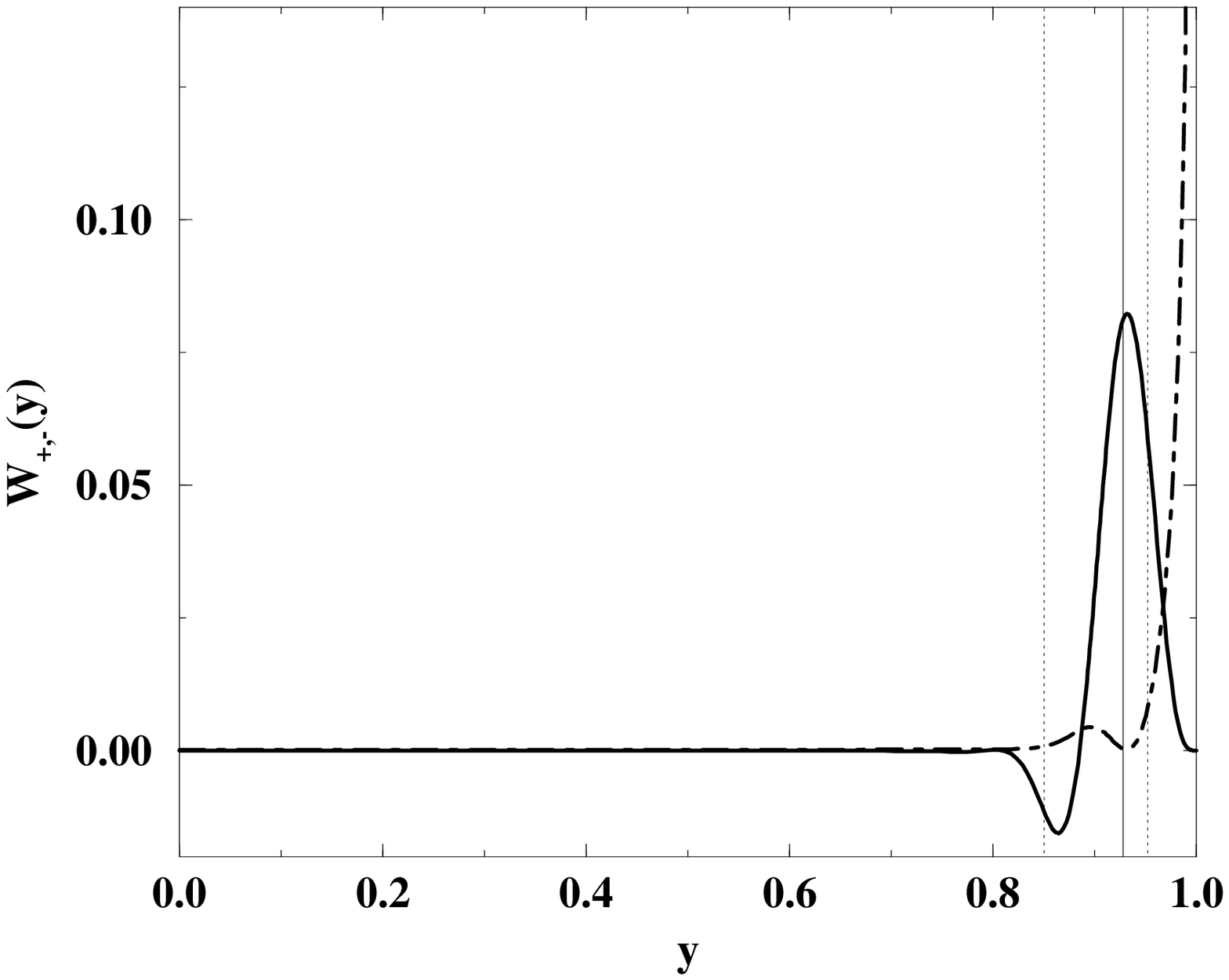} 
\caption{Energy balance for $m=0.9$, $p=0.85$. Energy 
roduction $W_+(y)$, solid line; dissipation $W_-(y)$, dot-dashed line.
Upper panel: stable flow at $\Re=5772$ 
(with $\Gamma_+= -0.0114$, $\Gamma_-=0.0122$).
Lower panel: Marginal flow at  $\Re=46400$ 
(with $\Gamma_+=\Gamma_-=0.0053$). }
 \label{f:near} 
\end{figure}

We now turn our attention to Fig.~\ref{f:near}, in which the main
point of this Letter is demonstrated.  Here, the Reynolds number is
the same as before, but the mixing layer has been moved close to the
critical layer. It is immediately obvious that the earlier balance is
destroyed. The counter-production peak in the mixed layer is much
larger than before, making the flow more stable.  The wavenumber used
is that at which the flow is least stable for the given Reynolds
number at this $p$.  For $m=0.9$, the threshold Reynolds number is
$46400$.  Fig. \ref{f:marg} shows the energy balances at marginal
stability - the picture is qualitatively the same here as at
$Re\approx 5772$ for the neat fluid.
The main factor determining the stability is the production, which is
driven by the phase change caused by the viscosity stratification.
The dissipation on the other hand depends only on Reynolds number and
does not respond disproportionately to changes in viscosity.  In neat
fluids, the term containing $U''(y)$ in (\ref{modOS}) is always of
higher order within the critical layer. However, with the introduction
of a viscosity gradient within the critical layer, the gradients of
the basic velocity profile will scale according to the mixed layer
coordinate $\xi$.  An analysis in the critical layer indicates that
for $q \le O(Re^{-1/3})$, the term containing $U''/(U-c)$ is now among
the most dominant. Any reasonable viscosity gradient of the right sign
will pick up this term, leading to vastly enhanced stability. Note the
dramatic effect in $U''$ in Fig. 3.

Indeed, in the light of this discussion we can expect that the large
effect of retardation of the instability would even increase if we
make the mixing layer thinner. This is indeed so. Nevertheless, one
cannot conclude that instability can be retarded at will, since other
disturbances, differing from the primary mode, become unstable first,
albeit at a much higher Reynolds number than the primary mode; when we
stabilize a given mode substantially, we should watch out for other
pre-existing/newly destabilized modes which may now be the least
stable.

Finally, we connect our findings to the phenomenon of drag reduction
in turbulent flows. Since the total dissipation can be computed just
from the knowledge of the velocity profile at the walls, any amount of
drag reduction must be reflected by a corresponding reduction of the
gradient at the walls.  Concurrently, the energy intake by the
fluctuations from the mean flow should reduce as well. Indeed, the
latter effect was measured in both experiments \cite{97THKN} and
simulations \cite{01DSBH,00Ang}.  The question is which is the chicken
and which is the egg. In our calculation we identified that the
reduction in production comes first.  From Figs. 4 and 5 (upper panel)
which are at the same value of Re we see that the dissipation does not
change at all when the mixing layer moves, but the production is
strongly affected. Of course, at steady state the velocity gradient at
the wall must adjust as shown in Fig.~5 (lower panel). We recognize
that in a turbulent flow there are a number of modes that interact,
but we propose that a similar mechanism operates for each mode at its
critical layer, where both elastic and viscous effect determine the
mean flow. In the present calculation we can consider all the putative
unstable modes, and conclude that with a viscosity gradient similar to
that seen in polymeric turbulent flows the threshold Re goes up five
times (to 31000).  We leave the confirmation of this prediction to
future experiments.

  {\bf Acknowledgments}. This work has been supported by the Israel
 Science Foundation.

\end{multicols} 
\end{document}